# Microstructure evolution and mechanical behavior of Fe-Mn-Al-C low-density steel upon aging

[1]Alexandros Banis, [2]Andrea Gomez, [1]Vitaliy Bliznuk, [3]Aniruddha Dutta, [2]Ilchat Sabirov, [1,4]Roumen H. Petrov

[1] Department of Electromechanical, Systems and Metal Engineering, Ghent University, Ghent, Belgium

[2] IMDEA Materials Institute, Getafe, Madrid, Spain

[3] ArcelorMittal Global R&D, Ghent, Belgium

[4] Department of Materials Science and Engineering, Delft University of Technology, Delft, The Netherlands

*Corresponding author: Alexandros.banis@ugent.be*

## Abstract

This study focuses on the microstructure's evolution upon different aging conditions of a high-strength low-density steel with a composition of Fe-28Mn-9Al-1C. The steel is hot rolled, subsequently quenched without any solution treatment, and then aged under different conditions. The microstructure of the samples was studied by means of Scanning Electron Microscopy, Electron Backscatter Diffraction, and Transmission Electron Microscopy. The aging treatment leads to the formation of an ordered face-centered cubic $L1_2$ phase named κ-carbide. This study aims to characterize the formation and growth of these κ-carbides qualitatively and quantitatively under different aging conditions. Then, an effort is made to relate the fraction and size of this phase with the tensile properties of the steel to determine the optimal aging conditions that will lead to a good combination of strength and ductility. It has been found that the κ-carbides start to form intragranularly through concentration fluctuations of aluminum and manganese inside the austenite grain. Then, with the process of spinodal decomposition, they grow in size coherently with the matrix. During this process, the strength and hardness of the steel increase while maintaining a relatively high elongation. The best combination of high strength and ductility was achieved at the aging condition of 8 h at 550 °C with an ultimate tensile strength up to 1157 MPa and total elongation of 51%. Increasing the aging temperature and time, κ-carbides start to form intergranularly, lose their coherency with the matrix and severely compromise the hardness and strength. The shearing of the carbides during deformation is also studied.

## 1. Introduction

Fe-Mn-Al-C steels are drawing the attention of the automotive industry lately due to their outstanding mechanical properties [1] and low weight. The weight reduction is attributed to density reduction by 1.3% for alloying with 1 wt.% of Al addition [2] by increasing the lattice parameter (for austenite: 0.00094 nm per wt.% Al) [3]. Chen et al. have extensively reviewed the current state of the art in low-density steels [4]. Depending on the chemical composition, these steels may be fully austenitic or duplex steels consisting of austenite, DO3 and B2 phases and α or δ ferrite [5–10]. This work focuses on fully austenitic low-density steels due to their better processability compared to the duplex ones [11]. Such microstructure is achieved by alloying with 5-12 wt. % Al, 12-30 wt. % Mn, and 0.6-2 wt. % C [12–14], and such steels are age-hardenable due to the precipitation-hardening effect of κ-carbides [15,16]. More specifically, according to the literature [2,17,18], grades containing 5-11 wt.% Al and 0.3-1.2 wt.% C are age-hardenable when heat-treated in the temperature range of 450-700 °C.

The κ-carbide has an $L1_2$-type structure and belongs to the space group Pm3m [19]. At aging temperatures between 450 – 650 °C, chemical modulation leads to homogeneous precipitation of κ-carbides coherent with the austenite matrix. The κ-carbide phase is metastable $(Fe,Mn)_3AlC_x$ (x<1) with a cube-on-cube orientation with the austenite matrix [20,21].

At higher aging temperatures (650 – 800 °C), intergranular κ-carbides nucleate heterogeneously on the γ/γ boundaries [22–24]. These intergranular κ-carbides are characterized by faster growth kinetics than the intragranular carbides and, consequently, a larger size [25]. Acselrad et al. [26,27] proposed an isothermal transformation diagram in which, for temperatures between 550 and 650 °C, exists a spinodal decomposition region at the initial stages of aging, followed by homogeneous precipitation of isomorphic κ-carbides within the austenite matrix, in which a regular periodic structure of the κ-phase can form. For longer aging times, heterogeneous precipitation on grain boundaries will occur. Park et al. [28] found that increasing the Mn content may delay the formation of intragranular κ-carbides, while the addition of C accelerates the formation of these carbides.

The formation of the κ-carbides leads to precipitation hardening of the steel caused by interaction between the κ-carbides and the mobile dislocations during deformation [29]. In addition, solid solution hardening plays an important role as a strengthening mechanism due to the high contents of C, Mn, and Al [30]. In some cases, such alloys may achieve exceptionally high elongations, ~100% [31], while they exhibit very high yield strength (YS) [32].

Recent studies have focused on the effect of additional alloying elements such as nickel [33–35], chromium [36–40], niobium [41], copper [42], sulfur [43], silicon [44], and vanadium [45]. Such elements may change the microstructure constituents and the transformation kinetics [46] and therefore have not been added in the alloy selected in this study. This work aims to characterize the precipitation mechanism of the κ-carbides from the early stages of aging in a high-strength low-density steel with a composition of Fe-28Mn-9Al-1C. By comprehending the formation of intragranular and intergranular κ-carbides, and their relationship with the austenitic matrix, it is possible to determine their size and fraction and evaluate their effect on the deformation mechanisms that take place during uniaxial tensile testing.

## 2. Experimental

The steel selected for this study has the composition shown in **Table 1**, in wt.%.

**Table 1**: The chemical composition of the studied alloy, in wt.%

| Fe | Mn | Al | C |
|---|---|---|---|
| balance | 28 | 8.7 | 0.9 |

Alloying with Al reduces the steel's density by 1.3% per 1% Al by expanding the lattice parameter. According to the density equation proposed by Frommeyer et al. [47], the density of the studied steel based on its composition is calculated at 6.576 g/cm$^3$, which is much lower than typical austenitic steel, which has a density of ~8.15 g/cm$^3$.

After casting, the ingots were reheated at 1200 °C for 8 h, followed by hot rolling until a total reduction of 65% in thickness was achieved. The rolling was performed in seven passes, and after the final pass, the sheet with a thickness of 5 mm was quenched in water to avoid the formation of coarse intergranular κ-carbides.

After the hot-rolling, rectangular samples were cut from the sheet, with the long axes parallel to the sheet rolling direction and were aged in a muffle furnace. To minimize decarburization, the samples were wrapped in stainless steel foil and submerged in preheated sand to eliminate the possible thermal gradients in the furnace as well as to ensure high heating rate. The studied samples are named after the aging treatment, and the conditions chosen for the aging treatment are described in **Table 2**. At the end of the aging treatment, the samples were removed from the furnace and the sand and were air-cooled.

**Table 2**: The aging conditions of the studied samples

| Sample name | Temperature (°C) | Time (h) |
|---|---|---|
| No aging (NA) | - | - |
| 550_1 | 550 | 1 |
| 550_2.5 | 550 | 2.5 |
| 550_8 | 550 | 8 |
| 650_8 | 650 | 8 |

The aging treatments were chosen based on the literature, hardness measurements, and the resulting microstructures. Aging temperatures as low as 550°C would require enough long aging time to achieve the peak aged conditions and even longer for the growth of intergranular κ-carbides. The selection of such low aging temperature allows easy to follow the microstructural changes (precipitation kinetics of κ-carbides) in the laboratory conditons and provide opportunities for better control of the aging time in industrial conditions. On the other hand, for higher aging temperatures, such as 650 °C, the peak condition is reached at very short times, less than 30 minutes. Therefore, it is more difficult to control the microstructure due to the faster kinetics for the growth of the carbides at 650 °C.

For Scanning Electron Microscopy (SEM) and Electron Back-Scatter Diffraction (EBSD) analyses, standard preparation procedures were followed. The samples were cut so that the entire thickness of the sheet could be observed from the transverse direction (TD). After mounting, they were ground with SiC grinding papers up to #2000 (SiC/in$^2$) and then polished with diamond pastes of 3 μm, 1 μm, and finally with 0.1 μm colloidal silica (OPU) for 20 min. For the SEM analysis, chemical etching was included in the process, using 10% $HNO_3$ in $C_2H_5OH$ (Nital 10%) for 2-5 s. The Transmission Electron Microscopy (TEM) samples were produced by grinding up to a thickness of 100 μm and using electrolytic preparation equipment with perchloric acid.

An FEI Quanta TM 450-FEG-SEM was used for the SEM and EBSD analysis. The voltage was 20 kV, with a final aperture size of 50 μm and a spot size number 5, which corresponds to a probe current of ~2.4 nA. The working distance was selected in the range of 10-15 μm. The imaging mode for SEM was back-scatter electron imaging (BSE). For EBSD analysis, the sample was tilted 70°. The resulting patterns were acquired on a hexagonal scan grid by a Hikari detector operated with EDAX TSL–OIM-Data Collection version 7.3 software. The scan size was 1000 x 1000 μm, and the step size was 0.6 μm. The corresponding orientation data were post-processed with EDAX-TSL-OIM-Data Analysis version 7.3.1. software using the following grain definition: misorientation with neighboring grains higher than 15°, a minimum number of points per grain was 8, and confidence index (CI) larger than 0.1. The raw EBSD data were post-processed (cleaned) in two steps to re-assign the dubiously indexed points using the *(i)* grain confidence index standardization and *(ii)* neighbor orientation correlation procedure, and the number of changed points did not exceed 0.5%.

A Jeol JEM-2200FS, 200 kV field emission transmission electron microscope was used for the TEM, Scanning TEM (STEM) and High-Resolution TEM (HRTEM) analysis. Due to the {001} cube-on-cube κ/γ orientation relationship, the κ-carbides were observed in the [001] zone axis [48].

For the Transmission Kikuchi Diffraction (TKD) analysis, the same TEM samples were placed on the specific holder with a tilt towards the EBSD detector of -30°. The working distance was 5 mm, and the step size used 6 nm. The aperture and spot sizes used were the same as those used for the EBSD analysis.

The Vickers hardness (HV) of the material was measured with a micro-hardness indenter with a load of 10 kg. Twelve indents were done per sample to acquire statistically representative results. For the tensile tests, sub-size samples of the ASTM E8 standard geometry, with gauge dimensions of 25 mm length by 5 mm thickness by 10 mm width were cut from the hot-rolled sheets parallel to the rolling direction. The strain rate for the tensile tests was $10^{-3}$ s$^{-1}$, and 3 samples per condition were tested.

## 3. Results

### 3.1. *Austenitic microstructure*

**Figure 1** shows the BSE-SEM images of the studied samples. All aged samples have a grain structure identical to the hot-rolled sample (**Figure 1a**), consisting of equiaxed austenite grains with annealing twins. In sample 550_8 (**Figure 1d**), fine precipitates marked with a yellow arrow are observed on triple junctions. In sample 650_8 (**Figure 1e**), these κ-carbides appear much coarser than in sample 550_8 and nucleate on the triple junctions and the grain boundaries (cf. **Figure 1d, e** yellow arrows). Specific large morphologies that may correspond to the intragranular κ-carbides are also observed within the austenitic grains, and they are marked with a red arrow in **Figure 1e**. The annealing twins do not appear to act as nucleation sites for the precipitation of intergranular κ-carbides, as no precipitates are observed on any of the twin boundaries.

The average number of grains taken into account in each EBSD scan is ~35000. The resulting grain size (diameter) is given in the diagram in **Figure 1f**. It can be observed that the grain size is not affected by the aging treatment, as it remains the same for every sample, with a minor deviation, at around 8 μm. This diagram also shows that a smaller fraction of grains reaches a diameter of 25 μm. These measurements account for the austenite grains and the annealing twins, as they are also represented by high-angle grain boundaries.

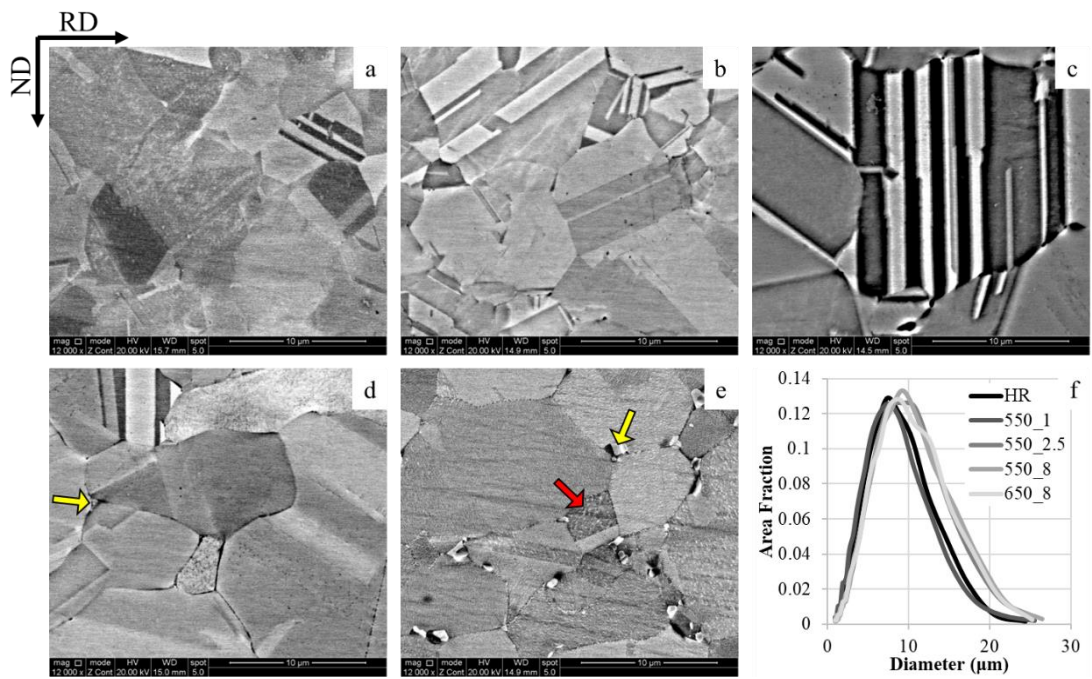

**Figure 1**: BSE-SEM images of the a) NA b) 550_1 c) 550_2.5 d) 550_8 e) 650_8 samples. f) Grain size (diameter) distribution obtained from the EBSD analysis.

*3.2. Precipitation and growth of intragranular κ-carbides*

TEM and STEM were utilized to observe the κ-carbides in more detail. **Figure 2a** shows an austenite grain and an annealing twin of the non-aged sample. There is no evidence of the formation of κ-carbides in this sample, which is attributed to the quenching immediately after the last hot-rolling pass. The precipitation of very fine intragranular carbides can be observed in the 550_8 sample in **Figure 2b**. At this aging stage, they appear to have fine size, making them difficult to distinguish.

On the other hand, increasing the aging temperature to 650 °C in the 650_8 sample, the κ-carbides become coarse with a well-defined cuboidal shape and form organized structures in the austenite matrix (**Figure 2c**). These structures consist of closely packed cuboids of approximately 15-20 nm, forming parallel stacks with broad γ-channels in the interspace, and this interspace can vary from 15-40 nm. This morphology is similar to that observed by Yao et al. [48], Acselrad et al. [49], and Gutierrez-Urrutia et al. [50]. This can be observed clearly in the dark-field (DF) TEM image of **Figure 3c** for sample 650_8, in which the cuboidal carbides and broad and narrow γ channels are shown.

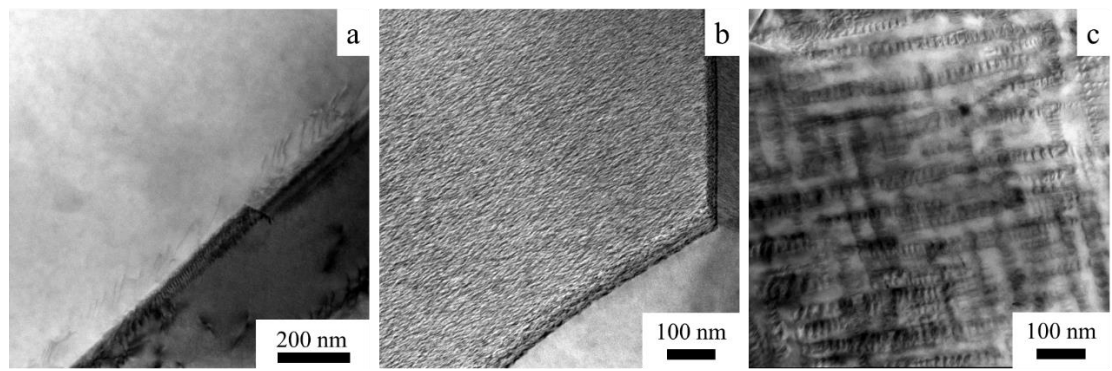

**Figure 2:** STEM images of the a) NA, b) 550_8, c) 650_8 samples.

This is not the case for the 550_8 sample (**Figure 3a**). In this image, the carbides have a globular shape, and they are much smaller and randomly arranged in the austenite matrix. The brighter areas indicate the start of the formation of parallel stacks and γ channels. From the dark field TEM images, it was calculated that the average size of the κ-carbides in the 550_8 sample is

~4 nm with a volume fraction of ~6%, while for the 650_8 sample, the average size of the carbides is ~20 nm, as mentioned before, and the volume fraction ~30%. The dark field images are taken from the [001] zone axis, as can be seen from the selected area electron diffraction (SAED) patterns of **Figure 3b, d**. The (010) and (110) superlattice reflections of the carbides are typical for ordered structures, confirming the cube-on-cube orientation relationship with the austenite matrix: $[100]_\kappa$ // $[100]_\gamma$ and $[010]_\kappa$ // $[010]_\gamma$, with a lattice misfit of less than 3% [20]. As reported, the spinodal decomposition and ordering are concomitant [51]. Diffraction satellites around diffraction and incident spots (see incert on Fig. 3b) originate from fluctuations of the C composition along the <100>, while the presence of $E2_1$ superlattice reflections is from the concurrent ordering of C and Al. The distance between the satellite centers and the diffraction spot centers let us estimate the concentration wavelength in decomposing solid solution, namely 5.5-6.0 nm. The (010) superlattice reflection was selected for both dark field images.

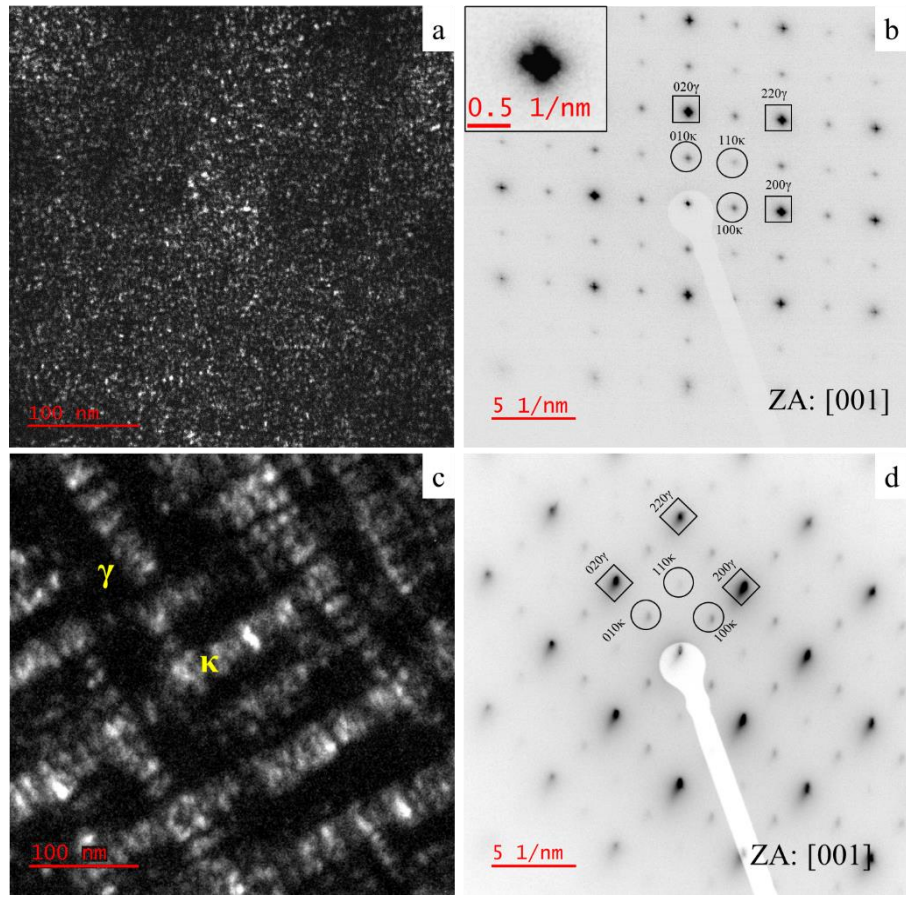

**Figure 3**: a) DF TEM image of the 550_8 sample showing the randomly dispersed κ-carbides. b) SAED pattern of the [001] zone axis from which the DF of image a was obtained. The magnified (200)γ diffraction spot with the diffraction satellites is superimposed on the image. c) DF TEM image of the 650_8 sample showing the organized κ-carbide stacks. d) SAED pattern of the [001] zone axis from which the DF of image **c)** was obtained. The DF images are taken from the $(010)_κ$ reflections.

The intragranular κ-carbides are formed through a spinodal decomposition that causes fluctuations of C and Al within the austenite [21,52]. Then, through short-range ordering, the Al-rich austenite areas transform into the κ-carbides after the further ordering of C atoms. Therefore, the κ-carbides contain a higher Al content than the γ matrix. This is confirmed in the STEM Energy-dispersive X-ray spectroscopy (EDX) analysis in **Figure 4**, where the content of Al in atomic % is shown. In the γ channels, the content of Al is lower, approximately 13 at.%, while in the carbide stacks, the content of Al increases up to 17 at.%. Atom Probe Tomography (APT) analysis [53] showed that the content of C follows the same trend as Al, with the matrix being depleted in C and the carbides rich in C.

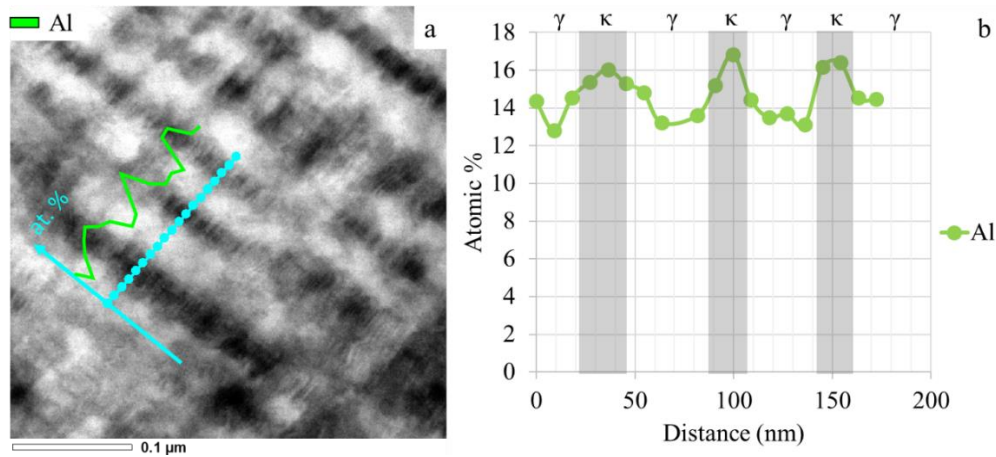

**Figure 4**: a) STEM EDX analysis on the κ-carbides of the 650_8 sample; b) the chemical analysis on the line scan allows for distinguishing the carbide from the matrix.

## *3.3. Mechanical properties*

The Vickers Hardness (HV) was measured for the studied samples, along with the tensile properties. The Vickers hardness data (grey circular points in **Figure 5a**) shows the same trend as the Ultimate Tensile Strength (UTS - black rectangular points) under different aging conditions. After aging at 550 °C, both HV and UTS increase with the aging time. For shorter times (1 - 2.5 h), both values increase faster; approximately an increase of 28 HV per hour of aging is calculated. For longer aging times, up to 8 h, this rate decreases, and the curve becomes flatter. In this part of the diagram, the speed of the hardness increase is calculated as approximately 7 HV/h of aging. Specifically, the NA sample has a hardness of 280 HV. Then, after 1 h of aging at 550 °C, the hardness increases to 390 HV. Thus, an increase of 100 HV in hardness occurs during the sample's aging for 8 h at 550 °C. Similarly, an increase in the UTS of ~100 MPa is observed for the same aging conditions, with a maximum of 1157 MPa for the 550_8 sample. Additionally, it is observed that the hardness of the 650_8 sample is lower than that of the 550_8. This trend is similar to that reported by [54,55], where it is claimed that the responsible for the first hardening stage is the spinodal decomposition, while for the second stage, the elastic and interface stresses generated by the growing κ-carbides are responsible.

**Figure 5b,c** present the engineering and true stress-strain curves for the samples aged at 550 °C, respectively. As was mentioned before, the UTS increases with increasing aging time by approximately 100 MPa. This increase is followed by a decrease in the total elongation (TE) of the material, which can be considered a measure of ductility. Specifically, the total elongation at fracture decreases from 77 % for the 550_1 sample to 49 % for the 550_8 sample. The yield strength (YS) is also increasing significantly for the sample aged for 1 h compared to the sample aged for 8 h, at ~ 200 MPa. The work hardening rate has been plotted as a function of the true strain in **Figure 5d**. In all tested samples, after the plastic strain of 0.03, the work hardening rate increases with further increasing plastic strain reaching the maximum value of 1931 MPa at the plastic strain of 0.29 in the 550_1 sample, 1755 MPa at the plastic strain of 0.26 in the 550_2.5 sample, and 1683 MPa at the plastic strain of 0.23 in the 550_8 sample. The work hardening rate tends to decrease with the increasing aging time, and the drop is more pronounced at the earlier stage of aging (from 1 h to 2.5 h).

The average values of the tensile properties for the three tested specimens per condition can be seen in **Table 3**, with the coefficient of variation (CV):

Table 3: The average values of the tensile properties for the studied samples.

| Sample | $\sigma_{YS}$ MPa | CV % | $\varepsilon_{YS}$ mm/mm | CV % | $\sigma_{UTS}$ MPa | CV % | $\varepsilon_{TE}$ mm/mm | CV % |
|---|---|---|---|---|---|---|---|---|
| **550_1** | 734 | 2.29 | 0.73 | 0.65 | 1019 | 0.31 | 0.77 | 1.27 |
| **550_2.5** | 926 | 0.56 | 0.77 | 3.12 | 1070 | 0.28 | 0.66 | 3.44 |
| **550_8** | 1015 | 0.44 | 0.86 | 0.62 | 1141 | 0.86 | 0.49 | 8.9 |

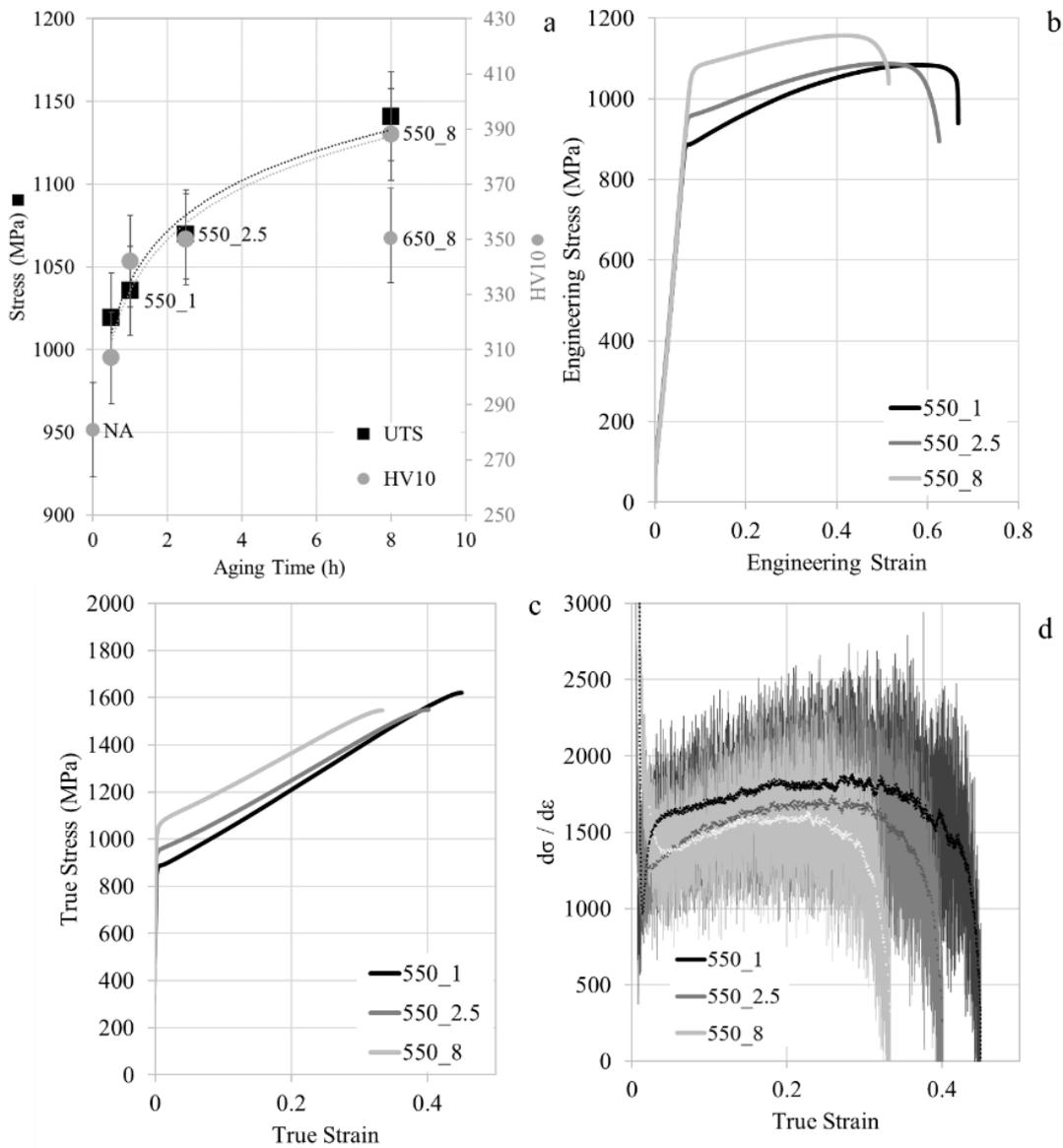

**Figure 5**: a) the UTS. and HV values for different aging conditions. b) Engineering stress-strain curves for the samples aged at 550 °C. c) The corresponding true stress-strain curves. d) work-hardening rate as a function of true strain with the trendlines overlapping the graph.

## 4. Discussion

As was shown from the SEM and EBSD analysis, the microstructure consists of equiaxed austenite grains with annealing twins. The grain size of the austenite grains does not change during aging due to the low temperatures used, and this implies that no recrystallization or grain growth occurs, and thus no new annealing twins are formed. Hence, the differences observed in the properties of the aged samples can be attributed only to the size and fraction of the κ-carbides. For the 550_8 sample, which exhibits the highest UTS and hardness, the carbides appear to have a globular shape and random distribution, as was observed in the DF TEM images in **Figure 3**.

Further analysis with HRTEM shows the arrangement of atoms between the κ-carbides and the γ matrix. **Figure 6a** is an HRTEM image that shows the atoms are oriented, and no different orientations are observed between the carbides and the matrix. Using the Fast Fourier Transformation (FFT) and masking the reflections of the austenite matrix, the image of **Figure 6b** is obtained. In this figure, the atoms that belong to the κ-carbides are depicted in white, while the austenite matrix is shown in black. There is no defined grain boundary between the κ-carbides and the austenite matrix; instead, the transition occurs gradually (blue areas). Also, it is shown that the orientation of the atoms does not change between the carbide and the matrix. In **Figure 6c**, a κ-carbide is shown in atomic resolution, in which the atoms can be seen, while the atoms in the austenite matrix are not so clearly visible. **Figure 6d** shows the exact carbide after the inverse FFT process, which shows no clear boundary between the carbide and the matrix.

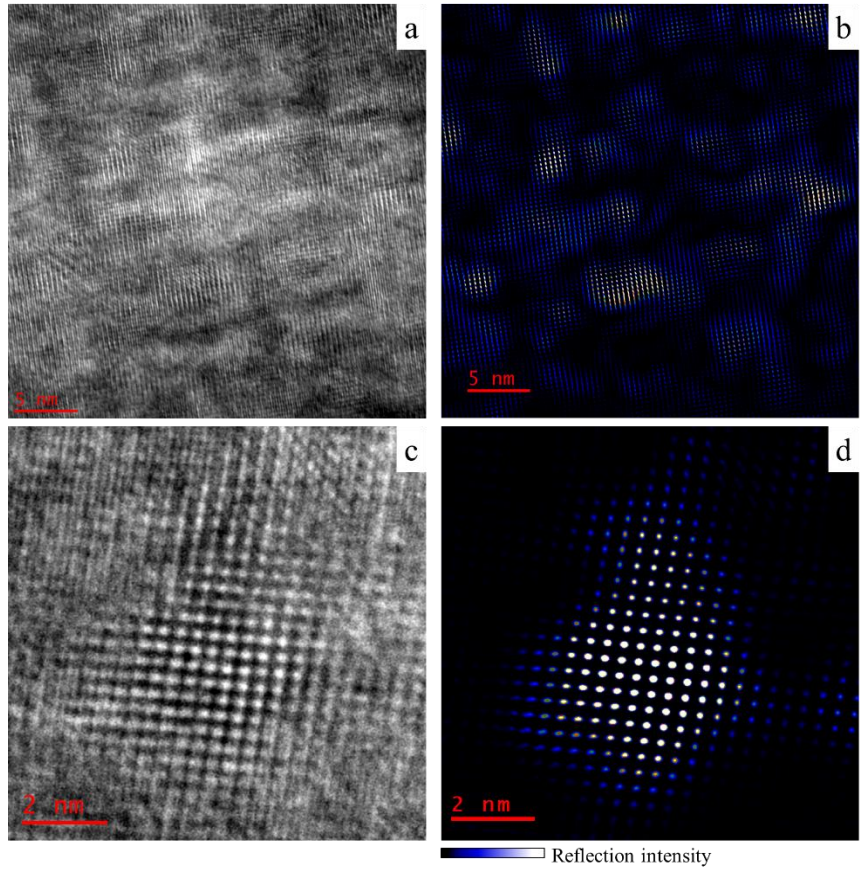

**Figure 6**: a) HRTEM of the 550_8 sample. b) the same area after the inverse FFT process, after masking the reflections of the austenite matrix. c) part of the same image showing only one intragranular κ-carbide. D) same carbide after the inverse FFT process.

Similar is the case for the 650_8 sample. Due to the coarser carbides in this sample, HRTEM could not give valid results, and TKD was used. In **Figure 7a** below, the Image Quality (IQ) map is given for the 650_8 TKD sample. The intragranular carbides are observed inside three austenite grains. When the different directions Inverse Pole Figure (IPF) maps are plotted (**Figure 7b-d**), these carbides cannot be revealed, indicating they have the same orientation as the austenite grains and are still coherent. This is expected as it has been shown that the intragranular κ-carbides remain coherent with the austenite matrix even at high aging temperatures [26] and longer aging times [48]. This is not the case for the intergranular κ-carbide located on the triple junction between the three austenite grains in **Figure 7b-d**. According to the literature [56], intergranular κ-carbides have a parallel orientation relationship with one of the neighboring grains. **Figure 7e** and **Figure 7f** show the point-to-point misorientations between the κ-carbide and Grain 1 (G1) and Grain 3 (G3), respectively, as measured on the dashed lines drawn in the figures. In both cases, the carbide and the two grains are separated by Low-Angle Grain Boundaries (LAGBs), with a maximum misorientation of 12°. Also, it is observed that the κ-carbide has a similar orientation with G1, but when rotated 90° around different axes, it has a similar orientation with G3. This is also confirmed by the Bright Field (BF) TEM image of **Figure 7g** and the diffraction patterns of **Figure 7h** and **Figure 7i**, which are taken from the κ-carbide and the γ grain, respectively. Both the grain and the carbide have the same [103] zone axis.

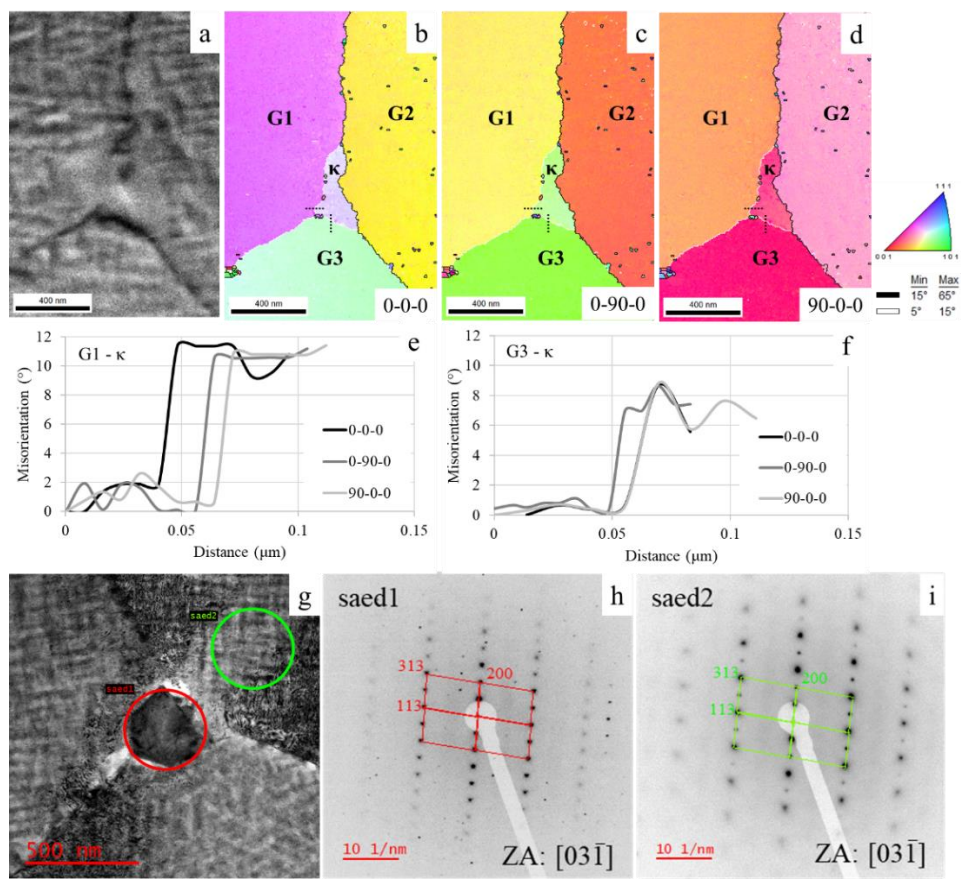

**Figure 7**: a) IQ- TKD map of the 650_8 sample. b) IPF map of the same scan, showing the orientations between the κ-carbides and the neighboring grains (G1, G2, G3). c) the same map rotated 90° towards the A2 axis. d) the same map rotated 90° towards the A1 axis. e) point-to-point misorientation between the κ and G1. f) point-to-point misorientation between the κ and G3. g) BF TEM image of a κ-carbide and its neighboring grains in sample 650_8. h) SAED pattern of the carbide marked in red. i) SAED pattern of the area of the grain marked in green

These intergranular κ-carbides are observed only on the 650_8 sample. The reason is the higher aging temperature, at which the Mn and Al start segregating on the grain boundaries. To confirm this, EDX analysis was performed in STEM. In sample NA (**Figure 8a**) and sample 550_8 (**Figure 8b**), the line scans for Mn (blue) and Al (green) show no peaks on the grain boundaries. This means that due to the quenching after the hot rolling, the Mn is not able to diffuse to the grain boundaries. Then, during aging at 550 °C, the temperature is still not high enough for long-range diffusion to the grain boundaries. Also, no segregation of Al or Mn is observed on the Twin Boundary (TB), and therefore no precipitation of κ-carbides is expected on these sites. In addition, it is well known that twin boundaries (Coincident Site Lattice - CSL) boundaries known as Σ3 CSL boundaries have lower energy than other high angle grain boundaries and therefore, precipitation, in this case, is impeded [57,58]. Only when the temperature is increased to 650 °C (**Figure 8c**) peaks for Al and Mn are observed. In this case, a coarse intergranular κ-carbide is shown. This carbide is located between two austenite grains and contains a high amount of Al and Mn, as shown from the EDX line scans. These intergranular precipitates are $L1_2$-type κ- carbides with a chemical composition close to the stoichiometric $(Fe,Mn)_3AlC$ [22,24,24,59]. It is also observed that these intergranular carbides are surrounded by a precipitation-free zone (PFZ). The presence of the PFZ in the 650_8 sample but not in the 550_8 sample indicates that the metastable intragranular κ-carbides have been dissolved, and the Mn and Al have been diffused towards the grain boundary. This difference in the chemical composition between the intergranular carbide and the PFZ attests that the lattice parameter is quite different; hence, the coherency between the carbide and the matrix has been lost.

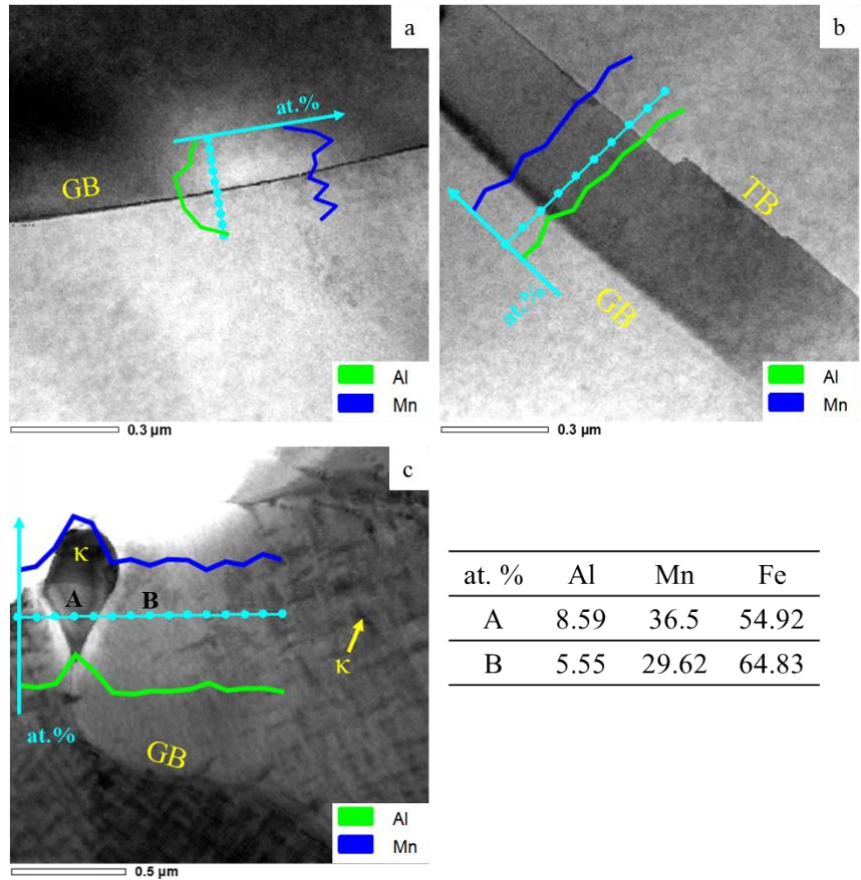

**Figure 8**: a) EDX on STEM on the grain boundary of the N.A. sample. b) EDX on STEM on the grain boundary of the 550_8 sample. c) EDX on STEM on the grain boundary of the 650_8 sample with a coarse intergranular carbide. The content of Al and Mn for points A (carbide) and B (PFZ) is also given.

Manganese has a larger atomic radius, 0.179 nm, than iron, 0.172, and therefore preferential substitution of Fe by Mn may increase the lattice parameter of the κ-carbide, thus contributing to an increase in strength and hardness during aging [46] due to stronger bonding between Mn-C couples in comparison to Fe-C pairs [60]. On the other hand, depletion of the PFZs from these elements leads to shrinking the γ lattice parameter and, thus, loss of coherency. These relatively soft PFZs are susceptible to plastic deformation, which leads to a reduction in yield strength [61,62]. In addition, the presence of incoherent carbides in the microstructure affects the mechanical behavior of the material negatively because they act as stress concentrators at boundaries [63]. Additionally, when the cube-on-cube orientation relationship is destroyed, coherency strain energy at the grain boundary increases. It is also reported that the grain boundary precipitations cause embrittlement and become a prior channel for crack initiation and propagation [25]. The above mechanism is also appropriate for explaining the lower hardness measured in **Figure 5a**. In addition, the difference in hardness between the 550_8 and 650_8 samples, can also be attributed to the solid solution strengthening mechanism, as the coarse, intergranular κ-carbides contain higher concentrations of Al, Mn, C. This leads to a depletion of the austenite matrix from these elements resulting in reduced contribution of solid solution hardening into the total strength, and thus to lower hardness values. These findings are in agreement with [64] and [25], while it is reported that the intergranular carbides have a similar effect on the ductility of the material [65].

In addition, Raabe et al. [15] and Gutierrez-Urrutia et al. [50] suggested that the predominant deformation mechanism is the Orowan bypassing of carbide stacks. Due to the narrow γ channels, dislocations cannot bypass the κ-carbides individually, but they can bypass the stacks containing the nanosized carbides. On the other hand, Yao et al. [48] described to a great extent the shearing of κ-carbides with an average size of 20 nm in steel with similar composition, aged at 600 °C for 24 h. Kim et al. reported that austenitic low-density steels containing κ-carbides deform in planar glide mode due to the slip plane softening originating from the shearing of κ-carbides by dislocation [65]. Additionally, due to the coherency between the intragranular κ-carbides and the austenite matrix, the looping of the carbides is not energetically favorable [62]. They suggested that the dominant deformation mechanism may change from the shearing of finer κ-carbides precipitation to the bypassing of coarser precipitates at a fixed volume fraction [66,67].

The deformation mechanism was studied in the 550_8 sample. In this sample, the size of the intragranular κ-carbides is relatively lower than that studied by Raabe [15] and Yao [48], ~4 nm, randomly dispersed and coherent with the austenite matrix. These samples exhibited the highest UTS of 1157 MPa while maintaining a total elongation at failure of 51%. EBSD analysis was performed on the deformed tensile sample at different strains. The EBSD scans shown in **Figure 9** have a size of 100 x 100 μm and a step size of 100 nm. In **Figure 9a-c**, the Normal Direction (ND) IPF maps are given, while in **Figure 9d-f**, the 2$^{nd}$ neighbor Kernel Average Misorientation (KAM) maps are shown. High Angle Grain Boundaries (HAGBs) are drawn in black, while LAGBs are in white. At the undeformed state (**Figure 9a,d**), the austenite grains can be seen with similar morphology to the SEM images of **Figure 1**. They are equiaxed, containing annealing twins, and the level of deformation given by the KAM map is respectively low. Then, at a strain of 29% (**Figure 9b,e**), the austenite grains are elongated towards the tensile axis, the annealing twins are still observable, and high levels of deformation are observed in the proximity of the grain and twin boundaries. Finally, the maps of **Figure 9c,f** are taken next to the fracture at a strain of 46%. In these maps, the austenite grains appear even more elongated; annealing twins are still present. Misorientations are observed inside the grains and twins, indicated by the gradients of color. No mechanical twins are present, which is attributed to the high stacking fault energy of the alloy, leading to a deformation mechanism of Micro-Band Induced Plasticity (MBIP) [68,69], while dislocations glide in a planar manner on the limited slip system without wavy and cross-slip [31,70]. Specifically, according to the literature and the SFE maps [71–74], the SFE of the alloy studied in this work is approximately 92 mJ/m$^2$. It has been well established that the acting deformation mechanism in alloys with such high SFE is the MBIP. Other mechanisms, such as Transformation Induced Plasticity (TRIP) and Twinning Induced Plasticity (TWIP), are active only in alloys with lower SFEs [71–74]. The KAM map shows high levels of deformation near the boundaries and inside the grains, while some unindexed areas appear black on triple junctions and the boundaries of smaller austenite grains. Elkot et al. also found that the initiation of the cracks occurs at these sites in H-charged samples [75]. As these sites are preferable for the formation of intergranular κ-carbides with PFZs, they pose possible weak spots in the material. The fractions of KAMs are given in the chart in **Figure 9g**. For the 0% strain sample, the KAM is very low, at 0.5°, at a fraction of ~55%. For the deformed samples, the peak shifts to the right, close to 1° and is broader than the peak of the undeformed sample, showing that a higher fraction of such misorientations are present. For the 46% strain sample, the peak is shifted more to the right, indicating that the level of deformation is higher. For the deformed samples, a second peak occurs at around 5°, indicating high levels of misorientations.

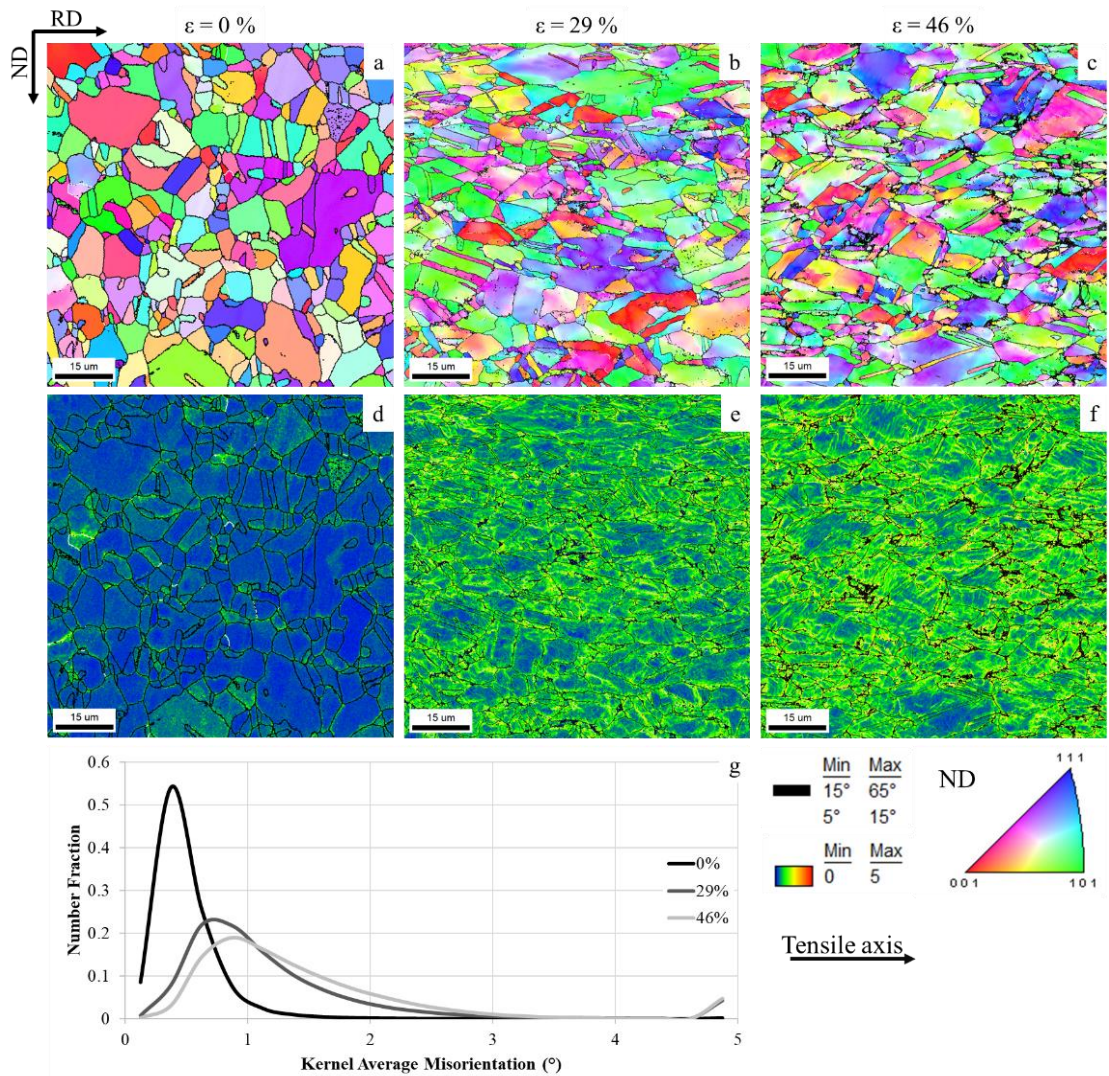

**Figure 9**: ND- IPF maps of the 550_8 sample at a) 0% strain, b) 29% strain, and c) 46% strain. d-f) 2$^{nd}$ neighbor KAM maps of the same scans. g) chart showing the fraction of KAMs for the 0, 29, and 46% strain samples

TKD analysis (**Figure 10a-c**) was performed with a step size of 6 nm in the deformed 550_8 sample at a strain of 36%. The tensile axis is parallel to the electron beam. In the IPF map of **Figure 10a**, an austenite grain is shown. In this grain, the misorientations caused by the deformation are visible in the gradient of the color. The misorientation from point to point is measured along the AB dashed line, and the result is shown in **Figure 10b**. The average misorientation is around 1°, but at a certain point, it reaches 6° and thus is recognized by the software as a low-angle grain boundary. The comparison of these results with the KAM map in **Figure 10c** shows that the misorientations are caused by the deformation inside the grain. This is also confirmed by the TEM analysis of the same sample, where the exact grain was found and studied. In this grain, parallel microbands can be seen (marked with yellow arrows). As can be seen from both the bright field and dark field images in **Figure 10e, f**, the carbides are diffracting only from the middle band of the image, from where the selected area electron diffraction pattern was taken. However, when tilting the sample 2°, then the carbides of the bottom band were visible in the dark field image. In order to study the shearing of the carbides, another grain was found in the same sample with a zone axis of [001], and an HRTEM image was taken (**Figure 10g**). Part of the image was taken, and using the inverse FFT, a κ-carbide was distinguished (**Figure 10h, i**) using the superlattice reflections of the diffraction pattern. The shearing of the carbide is visible (marked by the yellow arrows). As reported by Kim et al. [62], using in-situ TEM analysis in a sample with similar carbide size, the dislocations are gliding with frequent impingement, which indicates the shearing of the carbides. Thus, the carbide of **Figure 10i** has been sheared, most probably by the gliding of dislocations. When the fraction and size of the κ-carbides are higher, interparticle spacing is finer, and hence, the mobility of dislocations is reduced. The 550_8 sample consists of a higher fraction and larger size of κ-carbides compared to the 550_1 and 550_2.5 samples. Hence, this sample exhibits higher UTS and YS but decreased work-hardening. This decrease is attributed to the decrease of the slip band density and the softening effect caused by the shear of the κ-carbides [76]. The softening effect takes place as the leading dislocation gliding through the lattice meets the κ-carbide and destroys the SRO while shearing the carbide, and therefore, the following dislocations glide more easily, thus explaining the change in the slope of the tensile curves.

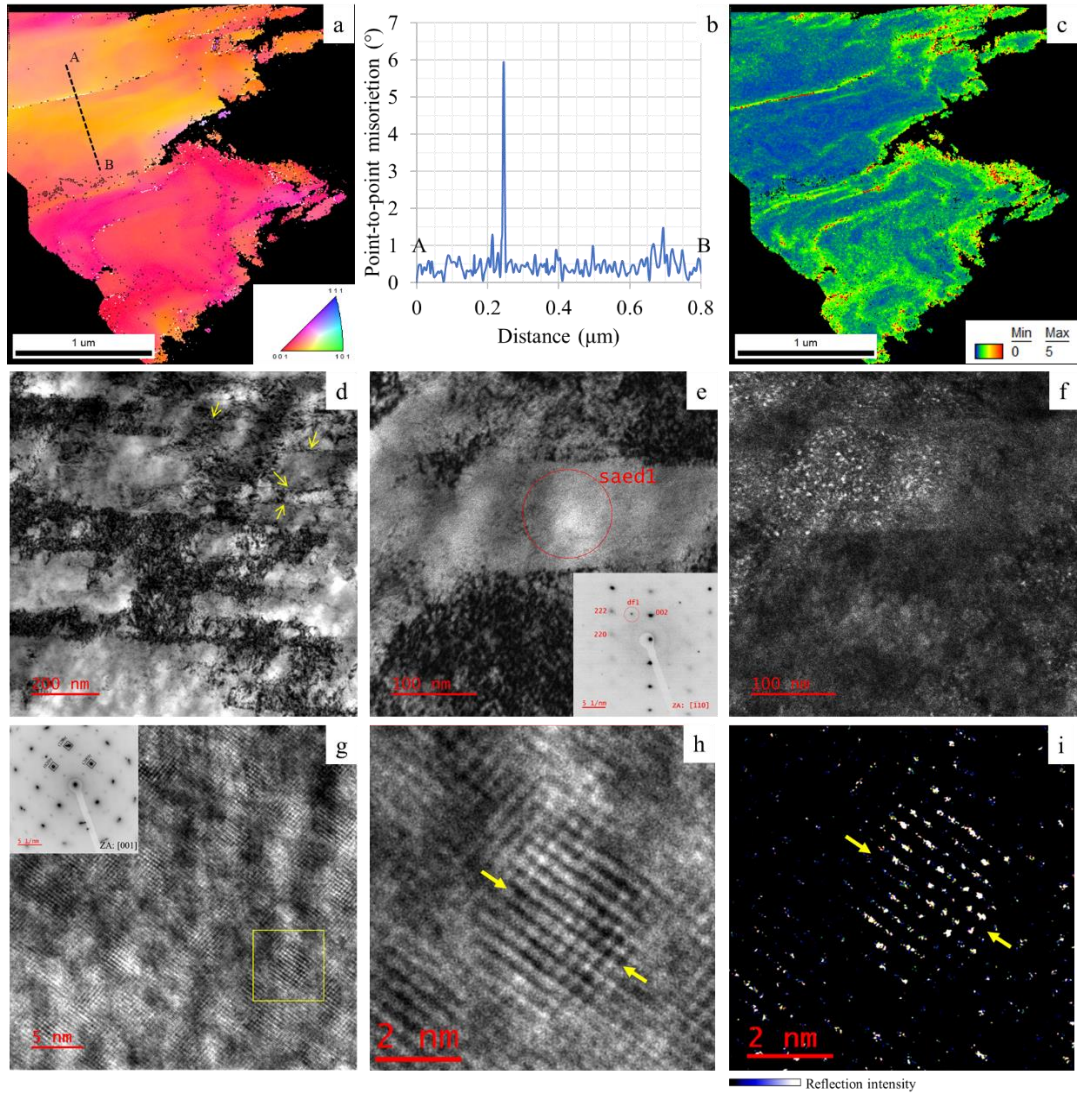

**Figure 10**: TKD IPF map of the 550_8 sample deformed at 36% strain. b) the point-to-point misorientation measured on the AB line of the IPF map. c) 2$^{nd}$ neighbor KAM map of the same grain. d) BF TEM of the exact grain showing different diffraction conditions and strain-induced bands marked with yellow arrows. e) part of image d showing the BF and the saed pattern from which the DF of image f) was taken. The carbides are diffracting from one band at a time, confirming the misorientation observed in the TKD g) HRTEM image of a grain in the same sample with a [001] zone axis. h) A κ-carbide shown in HRTEM and i) the same carbide after the inverse FFT process, indicating that it has been sheared.

## 5. Conclusions

This research studied the precipitation of intragranular and intergranular κ-carbides in a high-strength low-density steel with a composition of Fe-28Mn-9Al-1C during different aging conditions. The transformation phenomena and deformation mechanisms were studied by means of electron microscopy, and the following conclusions were drawn:

- The formation of intragranular κ-carbides via spinodal decomposition initiates at the samples aged at 550 °C, and after 8 h, they reach a size of 4 nm in diameter and a volume fraction of 6%. At this size and fraction, the material achieved the highest U.T.S., at 1157 MPa, while maintaining a ductility of around 49%. Further increase in the aging temperature leads to the formation of coarse intergranular κ-carbides, with high concentrations of Al and Mn. The higher content of these elements in the carbides expands the lattice parameter and leads to a loss of coherency with the matrix. Also, soft precipitation-free zones are formed around such carbides susceptible to plastic deformation.
- The intragranular carbides do not lose their coherency even at the over-aging condition of the 650_8 sample. Also, it was found that the intergranular κ-carbides retain the orientation relationship with the neighboring grains, as was found by TKD and TEM analysis.
- During deformation, the equiaxed austenite grains become elongated towards the tensile axis. Gliding and κ-carbide shearing are the main deformation mechanisms observed in this alloy. Slip bands cause misorientations in the grain, up to 6°, while the dislocations glide and shear the intragranular κ-carbides. Due to their coherency with the austenite matrix, dislocation bypassing is not possible for this fraction and size of carbides. Segregation of Mn occurs primarily in triple junctions, and the formation of intergranular carbides is favored in these sites, creating soft regions.

- The intragranular κ-carbides favor the properties of the alloy, as they increase its strength without compromising its ductility due to their ordered structure and coherency with the matrix. On the other hand, intergranular κ-carbides should be avoided, and the aging should be conducted at lower temperatures, at which the diffusion of Mn is impeded. Further studies will determine the threshold conditions to avoid grain boundary precipitation.

**Data availability**

The raw/processed data required to reproduce these findings cannot be shared at this time due to technical or time limitations.

**CRediT author statement**

**Alexandros Banis:** Conceptualization, Methodology, Validation, Formal analysis, Investigation, Writing – original draft, Visualization **Andrea Gomez:** Formal analysis, Investigation **Vitaliy Bliznuk:** Formal analysis, Investigation, Writing – Review & Editing **Aniruddha Dutta:** Resources **Ilchat Sabirov:** Supervision, Project administration, Writing – Review & Editing **Roumen H. Petrov:** Supervision, Project administration, Resources, Writing – Review & Editing

**Declaration of competing interests**

The authors declare that they have no known competing financial interests or personal relationships that could have appeared to influence the work reported in this paper.

**Acknowledgments**

The authors acknowledge the European Union's Research Fund for Coal and Steel via the DELIGHTED project (Grant No. 899332).